# Entangled responsibility: an analysis of citizen science communication and scientific citizenship

Niels Jørgen Gommesen

The Faculty of Business and Social Science, Digital Democracy Centre, Southern University of Denmark, nielsgom@sam.sdu.dk

Monday, 10th March, 2025

**Abstract**

The notion of citizen science is often referred to as the means of engaging public members in scientific research activities that can advance the reach and impact of technoscience. Despite this, few studies have addressed how human-machine collaborations in a citizen science context enable and constrain scientific citizenship and citizens' epistemic agencies and reconfigure science-citizen relations, including the process of citizens' engagement in scientific knowledge production. The following will address this gap by analysing the human and nonhuman material and discursive engagements in the citizen science project The Sound of Denmark. Doing so contributes to new knowledge on designing more responsible forms of citizen science engagement that advance civic agencies. Key findings emphasise that citizen science development can benefit from diverse fields such as participatory design research and feminist technoscience. Finally, the paper contributes to a broader debate on the formation of epistemic subjects, scientific citizenship, and responsible designing and evaluation of citizen science.

**Introduction**

The active involvement of volunteer contributors is critical for a successful citizen science project and its scientific outcome (Robinson et al., 2018). Citizen science (CS) depends on the public's meaningful involvement with science (Shirk & Bonney, 2018, p. 42). Thus, CS projects are often developed in ways that make it possible for the community of volunteers to participate in and contribute to multiple aspects of the scientific process, from the collection, selection, processing, and submission of data, over influencing research questions, to codesigning research methods (Novak et al., 2018), to communicating the project's research results and incorporating local knowledge. The benefits of a successful science-citizen collaboration can therefore be multiple. They may contribute to new opportunities, such as addressing local and national issues, contributing to scientific output, dialogical interactions with scientists, and potentially influencing policy and everyday matters of concern (Robinson et al., 2018, p. 29).

Citizen science connects scientific research with public engagement to potentially inform political decisions (Shirk & Bonney, 2018, p. 41). As citizens become involved in producing and communicating science, the entire research process and knowledge production are mediated, distributed, and shaped by citizens' engagements. Science communication is changing; it is no longer done by individual scientists but is increasingly produced in collaboration with the volunteer contributors (Hecker et al., 2018, p. 448). Schäfer and Kieslinger suggest that citizen science is comparable to a form of science communication (Schäfer and Kieslinger 2016). Hence, the citizen scientists are never mere data contributors their multiple intra-actions with digital technologies and artificial intelligence (AI) systems, communication officers and scientists, including their local communities, make them deeply entangled with the CS project, scientific knowledge production, including technological innovation (Shirk & Bonney, 2018). Irwin and Michael (2003), point out that because of humans' profound entanglement with technology, humans can be understood as hybrids that emerge from heterogeneous relationships between humans and nonhumans (2003, p. 133). Thus, according to this understanding, nonhumans are actively taking part in how scientific, political and ethical knowledge is produced (2003, p. 133); hence nonhumans have agency (Barad, 2007; Suchman, 2007) and take part in reconfiguring the scientific citizen (Alan, 2001). Irwin and Michal (2003) suggest that we consider how scientific citizenship is interwoven with technologies, environments, and animals (2003, p. 134). Hence, from this perspective agency and responsibility is extended and distributed throughout the technoscientific apparatus (Dickel 2020, p. 261) and can better be understood as the relational enactments of humans and nonhumans, and in terms of "what comes to matter" or are "excluded from mattering" (Barad, 2007).

Despite our knowledge about technoscience's implications in the modern knowledge society (Maasen et al., 2020) and its reconfiguration of the scientific citizen (Dickel, 2020, p. 262), few studies have addressed *how human-machine collaborations in a citizen science project and its technoscientific structures partake in reconfiguring the scientific citizen, science-citizen relations, including the very process of citizens' engagement in scientific knowledge production.* Stilgoe, Lock, and Wilsdon (2014) calls for new knowledge about these spaces for technoscientific engagement and their impact on scientific culture, politics and society (2014, p. 9-10). We need more knowledge about citizens engagement with new technologies and how their entangled practices shape, constrain and widens potentials for science and democracy (Nowotny, 2014). In light of recent innovations of digital CS technologies that integrate AI systems in the analysis and processing of data at unprecedent scales and speeds. There is an increased need for understanding human-machine collaborations in CS (Ponti & Seredko 2022; Seredko, Gander, and Ponti 2021) to understand how different human-machine entanglements and practices are reconfiguring and impacting citizen agencies and enactments of scientific citizenship. However, rather than assessing technology itself and keeping humans and technology apart, this paper will address the entanglements and agencies of humans and technology, drawing from Barad's agential realist theory (Barad, 2007), to create new knowledge on the technological apparatus of bodily production in The Sound of Denmark.

This paper addresses this gap by analysing citizens material and discursive engagement in the citizen science project The Sound of Denmark (Lyden af Danmark), a CS project developed at the Center for Macro Ecology, Evolution and Climate (CMEC) at the University of Copenhagen. This project collects large-scale data to identify sound sources, noise levels, and the distribution of natural and anthropogenic sounds in Denmark. The CS project invites citizens to contribute to sound mapping by recording, geolocating, processing and categorising sounds using their mobile phones and the mobile application (see fig. 1). The recorded sound data trains a machine-learning programme (AI system) that identifies sound sources. Participants can suggest new sounds to the AI system before finalising the contribution. Citizen involvement in the project contributes actively to scientific research and reshapes the epistemic tool for data collection (AI system), affecting the final data set and results. As a result, people become reshaped by the capabilities and limitations of technological infrastructure. Therefore, the CS project, the community, and the AI system comprise an interconnected technoscientific apparatus that iteratively restructures practices and agencies.

By doing this, it contributes to new knowledge on designing more responsible forms of citizen-science engagement that advance civic agencies. I will demonstrate how the citizen scientists are actively taking part in shaping the technoscientific processes in a citizen science context through their different intra-actions with the digital citizen science project The Sound of Denmark (SOD), and at the same time citizens are themselves shaped, constrained and empowered by the technoscientific structures.

In the following, I will outline the theoretical framework of this paper that builds on Karen Barad's Agential Realist ontology (2007). With this, *I expect to contribute new knowledge on how technoscience takes part in the reconfiguration of epistemic subjects, the enactment of scientific citizenship, and the bodily production of knowledge.* I discuss the methods and materials used in this study, followed by an analysis and discussion of the key findings.

**Agential realist framework for understanding the entanglements of which we are part**

According to Barad phenomena is co-constituted by material-discursive apparatuses of bodily production that are made of specific human and nonhuman intra-actions (2007, p. 206). Phenomena, she argues, are the effect of boundary-drawing practices that make some concepts, identities or attributes come to matter to the exclusion of others (2007, p. 208). She writes:

> The identities or attributes that are *determinate do not represent inherent properties of subjects or objects*. Subjects and objects do not pre-exist as such but are constituted through, within, and as part of *particular practices*. The objective referents for identities or attributes are the *phenomena constituted* through the i*ntra-action of multiple apparatuses* (Barad, 2007, p. 208).

Therefore, the phenomena including the inherent properties of objects and subjects do not pre-exist per se, but are constituted through specific material-discursive practices within and as part of apparatuses of bodily production. Following from this agential realist understanding materialisation is better understood as the intra-active engagements of humans and nonhumans. Barad (2007) that the notion of intra-actions is a key element in her agential realist framework and implies the mutual constitution of entangled agencies (Barad, 2007, p. 33). This is in contrast to the concept of interaction that assumes individual agencies precede their interactions, in contrast Barad's use of intra-action recognises that distinct agencies do not pre-exist but rather emerge through intra-action (2007, p. 33).

Hence from this conceptualisation agencies do not pre-exist as individual elements, it is not something that someone has (2007, p. 172) but exists only through their mutual entanglement in, e.g. a technoscientific apparatus.

Agency, according to Barad, is a matter of intra-acting; it is a doing, it is relational, and an enactment:

> Agency is about changing possibilities of change entailed in reconfiguring material-discursive apparatuses of bodily production, including the boundary articulations and exclusions marked by those practices in enacting a causal structure. Particular possibilities for (intra-)acting exist at every moment, and these changing possibilities entail an ethical obligation to intra-act responsibly in the world's becoming, to contest and rework what matters and what is excluded from mattering (Barad, 2007, p. 178).

As Barad points out the researcher or developer, the project team, who takes part in the enactments that shapes and constitutes the citizen science project, that is the technoscientific apparatus, are not without responsibility, because she/he/they are implicated in the production of the phenomena, and the bodily production of scientific knowledge (Juelskjær et al., 2021, p. 143). This includes the enactments that enable some concepts and knowledge to matter, excluding others. Hence, as Barad points out in the example because of our entanglements with the world and the apparatus in questions we are responsible for the knowledge we seek and partake in producing and the "consequences of our research" (Juelskjær 2021, p. 143) because it is a result of the practises, we have a role in shaping (2007, p. 203).

Hence, from an agential realist understanding, responsibility as researcher and codeveloper of citizen science entails ongoing responsiveness or response-ability to the entanglements of which they are a part (Barad 2007, p. 394). In agential realism, the researcher is an emergent part of the technoscientific apparatus and its human and nonhuman intra-actions. In so far as these research and participatory apparatus enact agential cuts that make some things come to matter to the exclusion of others. This implies that both the scientists, the citizen scientist, the technoscientific apparatus and the phenomenon are mutually reconfigured through the research and development process (Juelskjær et al., 2021).

This agential realist production of knowledge implies that to intra-act responsibly with and as part of the development of a CS project will require a critical exploration of the boundaries, constraints, and exclusions that we partake in enacting, and which operate through the intra-actions of the technoscientific apparatus. Therefore, by acknowledging that 'we' are an entangled part of the world's ongoing becoming together with human and nonhuman others, we realise that "the subject of knowing is not an individual but is linked to the research apparatus" (Juelskjær et. al 2021, p. 154).

Knowing and responsibility emerge from the intra-active engagements of the scientists, project/research team, citizen scientists, and the digital platform itself. Responsibility is relational and entails accountability for the material-discursive practices we partake in shaping because these entanglements reconfigure our very beings, our imaginations, and societies (2007, p. 383). The agential shift in theoretical perspective is crucial because it creates new opportunities for redefining unjust epistemic exclusions and constraints of the scientific citizen, and for taking responsibility for the designs we create, as well as to create knowledge about how they shape our very beings, doings, and knowledge. The next chapter outlines the methodological approach of this study.

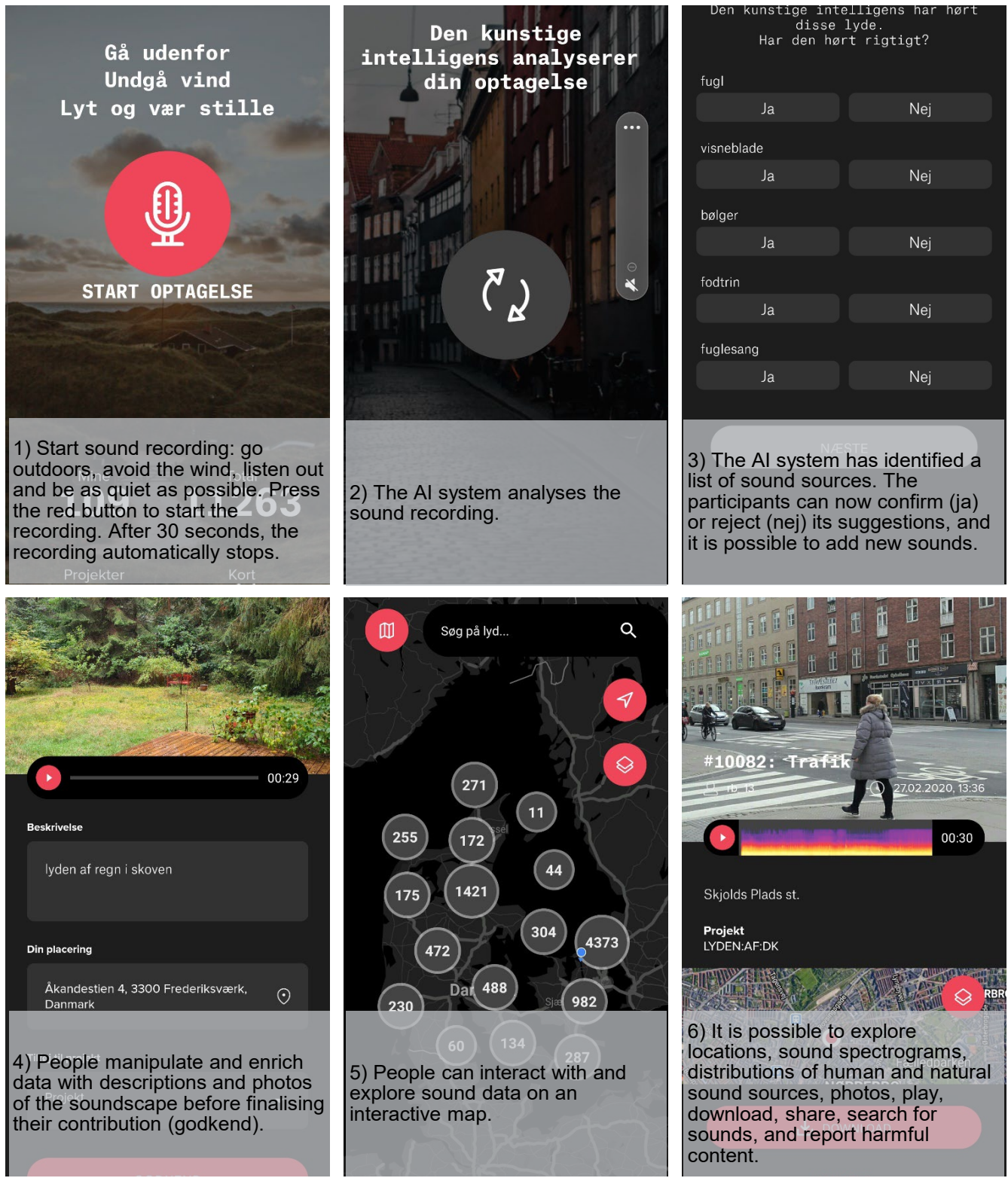

Figure 1. The mobile interface of the digital citizen science project The Sound of Denmark.

**Methods and materials**

*Empirical setting and role as codeveloper*

My empirical focus in this paper involves my ethnographic research in the citizen science project the Sound of Denmark, a research project developed in collaboration with the Center for Macro Ecology Evolution and Climate and the University of Copenhagen. As a member of the project/developer team, I spent 16 months, from March 2018 until June 2020, as an ethnographic and participatory design researcher (Blomberg & Karasti, 2012) focusing on how to design citizen science communication that strengthens citizens' democratic actions, through the co-development of a mobile app and digital citizen science project (fig. 1) in Participatory design collaborations with prospective citizen scientists and with the SOD project team and external developers. This role allowed me to act as a link between the participants and the developer/ project team (see Hughes et al., 1992). The participatory design approach in combination with ethnographic methods provided the opportunity to study the practices and contexts of prospective users/citizen scientists in a co-design context (Blomberg & Karasti, 2012), and as codeveloper and member of the SOD project team, to explore and experiment as part of the material-discursive practises that give shape to the digital development process.

My analysis of peoples' engagement across the SOD communication platform involved a focus on the intra-active engagements of the community of volunteers, and their exchanges and involvement across SOD's communication platform, that is: the SOD mobile app forum (fig.1), email exchanges between communication officers and citizens, and the Facebook group of the project, and a data corpus that includes (160) emails, (40) Facebook posts, and (30) threads from the mobile app's community forum. The documentary materials include text, photos and video materials (e.g. Facebook posts), including photographic materials of my interactions with the mobile app, and excerpts from the diary notes of a dedicated citizen scientist.

To understand the size of the SOD community, there have been 5745 contributions by volunteer contributors comprising over 11263 unique sound recordings and 2254 distinct sound sources found in the Danish soundscape. Among SOD's demographics are diverse ages, genders, etiquette and occupations, with participant groups like families, daycare children, elementary and secondary school children, pensioners, unemployed and academics. These groups have similar demographics to those of my earlier co-design workshops and interviews with amateur scientists and non-experts, where the community members are predominantly white and middle-class.

Based on the mail correspondences between the volunteer community and the communication officers, as well as my ongoing correspondence with a citizen scientist from the SOD community, I was able to map the group's demographics and examine their experiences as volunteers. However, it is not clear whether underserved groups are represented in the SOD profile database. Anonymized data has been used, and pseudonyms have been employed to protect the identity of those involved in the project, especially those who were not informed or had not provided consent. Moreover, since the excerpts of the data I used do not reveal the identity of the participants, I did not obtain consent from each individual who participated in the discussion across mobile forums, Facebook groups, and emails.

*Data collection*

I adapted a netnographic research approach inspired by (Kozinets, 2015; Kozinets, 2017) combined with virtual ethnography (Hine, 2000) as a general approach to studying the technoscientific apparatus of the SOD and the intra-active engagements and communication of the community of volunteers. My approach involved participant observations through instances of digital communication with participants, project members and developers across SOD's communication channels. I examined the intra-active and communicative engagements between the CS project team and the community of volunteer contributors, as well as how this communication developed across the project's communication channels: a community forum, a Facebook group, and correspondence with participants via email. Through their communication with the project, the volunteers provided information about their agencies, epistemic practices, experimental designs, social relationships, and reasoning styles (Mahr, 2021, p. 38). Through my ethnographic research, I explored, identified, and made visible the material-discursive practices of the technoscientific apparatus, its bodily production of knowledge, and its reconfiguration of the scientific citizen (Alan, 2001). In addition, my analysis of the citizen scientist Louise's diary notes allowed me to detail my understanding of the participants' experiences as part of the SOD community.

*Data analysis and thematic coding*

The data analysis focused on significant events in intra-active engagements and communications across the SOD communication platform, including diary notes of one of the citizen scientists. This revealed to me how the technoscientific apparatus reconfigured epistemic agencies and enacted scientific citizenship. My data analysis followed the recommendations of Braun & Clarke (Braun & Clarke, 2006) for inductive thematic analysis, which emphasise an organic approach that allows theming across the entire dataset. Here the researcher plays an active role in creating new themes and codes, and where the data suggests each theme's name. Hence, the researcher, in his work with empirical materials actively enact new entanglements (Juelskjær et al., 2021, p. 146). Furthermore, I use coded data grouped under recurrent themes in my empirical analysis. I use direct quotes from the data corpus for each theme. Coding involved: 1) familiarising myself with the data to establish classifications; 2) creating and collecting codes for each category through open coding in NVivo. 3) Identifying potential main themes and writing memos about them. This led to a more closed coding of observations and a reduction in the number of codes (Charmaz, 2006); categorisation and reduction of codes, 4) comparing codes for each theme. 5) Forming a final definition of themes, and 6) developing a preliminary analysis based on examples.

**Analysis and discussion:**

In the following, I present the empirical findings of my analysis of citizen science communication. I aim to analyse the material and discursive practices that operate through the SOD mobile platform and, by doing this, to create new knowledge and understandings of how they partake in reconfiguring citizens' capacities to enact citizenship and bodily production of knowledge. The themes in this chapter are analysed as follows:

- Theme 1: The citizen science project as a material-discursive apparatus of bodily production
- Theme 2: Enacting Scientific citizenship – entanglements as relations of obligation
- Theme 3: Entangled responsibility and enactments of concern
- Theme 4: Machinic agencies, boundaries and epistemic exclusions

**Theme 1: The citizen science project as a material-discursive apparatus of bodily production**

*The Sound of Denmark as technoscientific apparatus of bodily production*

This section analyses SOD as a technoscientific and material-discursive apparatus of bodily production by focusing on 1) the entanglements of the CS project's mobile platform, its AI system, and its community of volunteer contributors, 2) how they enact human-machine boundaries and relations, 3) to create new knowledge on how human-machine collaborations intra-actively reconfigure the scientific object (sound) in the study (Barad 2007, p. 383), epistemic practices, subjects in the community, and the production of scientific knowledge. In addition, human-machine interactions enact "cuts," which determine which concepts are realised and which ones are excluded; they evolve what is possible and impossible (2007, p. 234).

Citizens' material and epistemic practices involve many intra-actions with the mobile platform. For example, they are enacting active listening by recording city sounds (see figure 1), adding datasets, receiving and replying to feedback from AI systems, suggesting new sounds, uploading sound data, sound mapping, classifying, tagging, selecting, downloading, deleting, relocating, and sharing sound on the project's interactive map. But also reflecting critically on sound quality, sharing and discussing sounds with the project community, scientists and communication officers.

These entangled and epistemic practices are exemplified in Louise's (Pensioner citizen scientist) diary on Christmas Eve 2019:

> As a tradition, my partner and I walk on Christmas Eve while everyone is having dinner. We usually spend Christmas Eve in a dark forest with few sounds to record. Our family decided this year to walk to the Frederiksberg Alle Metro Station after celebrating our Christmas with some friends who live next to the new metro line after observing (listening) how others celebrate Christmas. During this time, we also made sound recordings of fun sound effects, such as a voice shouting out through an open window: "How many want port wine?". We took the metro to Nørrebro, got off the train, recorded sounds and photos, and went down to the next metro station (Skjolds Plads) and then up to a new place. This led to 14 audio recordings of Christmas Eve with photographs described in the report I sent to the project.

This example demonstrates that participants' epistemic practices are a part of enacting and reconfiguring what concepts are given a definition, and which one become excluded through their data collection and contribution to SOD. When the participants are recording sounds with the mobile app, they decide what to observe (people celebrating Christmas Eve), they decide what sounds to record, the location of their recordings, they document their findings with sound and photos and share their results with the project or the community. Furthermore, the results demonstrate, that their active engagement with the project is a social endeavour that is closely connected to their everyday life, something Louise and her partner do together to explore the city soundscape. Finally, it demonstrates that volunteer contributors are not only entangled in the production of scientific knowledge and other practices of knowing through their human-machine embodiments in the project (Barad, 2007, p. 379), by drawing attention to issues they care about that might be ignored by the scientist (Davies & Horst, 2016, p. 192), they are actively performing scientific citizenship.

The participants in SOD, intra-act as part of a larger material configuration that enact boundaries and exclusions of epistemic practices. As demonstrated in a mail from Beth:

> Hi, I think it is a fantastic idea (project), and I would very much like to be part of it. I have already recorded some sounds. In most occasions, unfortunately, *it will not play the recorded sound; hence I have to write tags from memory*. This means that I might emphasise something that does not sound well in the recording. However, I cannot continue without saving and can no longer add new tags. Instead, I suggest you change the setup and make it possible to add tags for 10 minutes after a recording, so you can listen to the sound after it has been uploaded and then add the extra sounds.

This example demonstrates that the mobile app's technological issues are delimiting the participants' epistemic practices and agencies. First, the participants cannot relisten to their sound recordings. Therefore, they cannot process and analyse their recordings accordingly. Finally, they cannot correctly tag the sounds and enrich the data, which means that the AI system is receiving flawed data from the users. The experience is unsatisfactory for the participants, and it affects the quality of the uploaded sound recording and, therefore, the training of the AI system.

The examples demonstrate the material-discursive practices of the apparatus and how the intra-actions of humans and nonhumans define what concepts come to matter and which ones are excluded (Barad, 2007, p.234) from the data collection process. The participants and the AI system are intra-acting with one another. On one side, participants reconfigure the technological environment of SOD; on the other, people are shaped and reconfigured by the system. Citizens reconfigure the AI system through sound recordings, suggestions of new sounds, and corrections of the AI feedback when they are using microphone devices to enhance the sound quality of their recordings. Moreover, technological issues in the app restrict the participants' epistemic practices and reduce their opportunities to utilise and share epistemic resources in the community (Dotson, 2012).

Another example that demonstrates how epistemic constraints and exclusions operate through SOD is issues with poor sound quality and inadequate sound recognition. As Henry tells us in a thread in the community forum:

> Why is the sound (quality) so bad? I have recorded excellent sounds with my phone, but they are barely recognisable in SOD. So far, I have listened to around 40 recordings; the best is one eating an apple, bird song sounds horrible, and my recording of a street orchestra at Nørrebro in Copenhagen is highly distorted. What is the point?

Henry's sound explorations demonstrate that the SOD mobile app distorts the sound quality in his recordings. His comment suggests that participation is pointless because of the poor sound quality. This is consistent with other responses from the community. An anonymous participant writes: "bad sound! When I listen to my recordings it sounds like it is recorded through an old can", and Louise: "I have recorded quite a few audio files but also deleted some again because they were too bad". These examples suggest that people are thinking critically about the poor quality of sound in the project. The poor sound quality is frustrating to most participants and discourages them from continuing to participate. The examples demonstrate that the technological environment is enacting boundaries and constraints that makes it difficult to contribute with high quality sound recordings. In some circumstances this can lead to anti-programmatic behaviour in the community with people following alternative programs and practices that diverge from the project protocol and scientific focus.

*Machinic reworkings, constraints and exclusions*

A number of community members have pointed out issues about inadequate sound recognition from the AI system. Louise (pensioner, citizen scientist) describes how it affects her engagement with the project:

> Once the AI has analysed the recording, it makes different sound suggestions. However, there are regular descriptions of sounds that I do not recognise, so I have to use the no-button, even if this means that I do not recognise the sound. Many of these sounds are unknown to me and cartoonish sounds. I cannot recall them, but I miss an sound dictionary where I can listen to the ML system's unusual sound suggestions.

The community of volunteers experienced incorrect sound recognition feedback from the AI system based on their intra-actions with the app. Together with the poor sound quality, it caused confusion for several participants. It not only constrained the quality of their contribution but also limited their possibilities for being fully engaged with the project's epistemic resources (Dotson, 2014).

Louise's description of her sonic explorations in the forest, demonstrate her critical thinking and concern about the limitations that operate through the project, and how epistemic practices are constantly reworked:

> I tried to record more sounds with the app. It works best when recording offline. It takes way too long with the artificial analysis. I am losing my patience. It is easier to add the description of the sound yourself. However, maybe that means the robot is not getting smarter? There is a conflict between user and project interests, and I do not like the long waiting time – after all, the AI cannot provide a reasonable description of the sounds.

The example demonstrates how her engagement with the mobile platform and AI system reconfigured her epistemic practices and capacities to act. Louise found the app works best offline because the AI algorithm spends a lot of time processing data. Since the system's data analysis is slow and its sound recognition is inadequate, she found it easier to add sound descriptions manually. The shift from automatic to manual sound description, a reconfiguration of practice, makes her think about how it will impact AI system training.

**Theme 2: Enacting Scientific citizenship – entanglements as relations of obligation**

This theme demonstrates that citizen scientists in SOD are more than volunteers collecting and contributing to scientific research. Through material-discursive entanglements with SOD's technological milieu, they bring attention to different matters of concern and interests that affect the continued development of the project and how knowledge and epistemic subjects are intra-actively produced. The community of volunteer contributors enact themselves as citizens through their intra-active engagement with the SOD mobile platform and ML system, including data contributions to the production and transformation of scientific knowledge production.

*Citizen science as social and scientific practices*

Key findings from my early interviews with citizen scientists reveal that epistemic practices are often performed differently through their active engagement with a project. Inge (age 45, veterinarian) describes her participation in the CS project, The Ant Hunt as a social and collective experience with her son:

> I have an 11-year-old boy who thinks natural engineering is great fun, so that was actually the background. During our summer holidays, we found The Ant Hunt as something we could do together, and I think it was an interesting study to do and a good opportunity to teach him something about research.

Their material agencies with this project express their mutual engagement, involving social and scientific practices, and a means by which they can both discuss and share something that interests them. Their material-discursive practice with the project enables them to explore their curiosities and gain new knowledge about ants. Inge explains: "We want to find out what kind of ant we have down there (in the ground) because then you can read more about it". Together, they create a dialogical space where they can collectively exchange ideas and process their uncertainties. Inge explains: "when we have been walking in the woods, we have talked about ants that live in different ways, look different, and things like that". Hence, Inge and her son's active engagement with the project in their local environment is also a learning process about science, caring for the natural world, and the ant species they discover. Inge states: "in general, it is about teaching one's child to take an interest in nature and observe and put into words what we see". Their entangled practices with the CS project, local environment, and the ants enhance collective learning and help formulate new understandings and knowledge about the world.

Inge explains: "You learn about methods and how to register, all the things they are not learning in school – 'scientific methods' – I think it is exciting to get more knowledge about where you are [local environment]". Taken together, their different intra-active engagements with the project materialised through epistemic practices and dialogic communication become the means to enact scientific citizenship that enhances their epistemic and democratic capacities to act on their interests and contribute to the very shaping of scientific knowledge production in the project.

*Citizen scientists take ownership of the project*

Key findings suggest that citizens want to participate on their terms by pursuing their own interests, values, and passions. Moreover, they wish to interact with others in the community and share their experiences. In a mail to SOD, Hanne (kindergarten teacher) describes her experiences and participation in SOD with children 3-4 years of age:

> Over a five-week period, we integrated SOD into our projects to focus on the sound of the water; we discovered that every city fountain made different sounds. At the children's requests, we threw stones of various sizes in Limfjorden and listened to the different sounds. At Vilsted lake, we repeated the session - it sounded like music. Back in the daycare, we addressed different words for water, wrote them down, and painted our experiences with watercolours. Based on our experiences with sound, we prepared small scientific experiments with water and talked about the importance of caring for the water.

The example demonstrates how the daycare through diverse material-discursive practices took ownership of the CS project by drawing attention to issues they care about and by integrating them into their everyday lives. Their intra-actions with the project set a different direction that opened for further involvement and collective ways to think about and with sound. Their intra-actions with the project, explorations of soundscapes and places in the city, experiments with listening and sound-making, scientific experiments, and playful experiments with words and language describing water. Demonstrate how the citizens not only contribute with new layers of meaning, value, curiosity, and imagination to a CS project but also how different material-discursive "practices of citizenship" can create and enhance civic capacities (see Horst 2016, p.195).

    The citizen scientists' relational ways of thinking and doing with the project, other people, things and natural environment demonstrate their diverse forms of epistemic practices and capacities to restage things in new situations and contexts, hold great potential for generating alternative forms of listening practices and engagement beyond the focus of the scientists, to strengthen civic capacities, and cultivate new potential sites of communication that enhance scientific citizenship.

*When citizen scientists criticise and contest the scientific process*

The mail exchanges between Louise (pensioner and dedicated citizen scientist) and the communication officers in SOD, together with descriptions from her diary about personal experience, thoughts, practices and sound experiments form the project, and her concern about the technological issues and shortcomings she discovered along her engagement with the project, reveal that citizens' diverse epistemic practices are part of and contribute to the project's data collection, it is not limited to the critique of power but recreate their relations intra-actively through that critique.

Louise writes: "I have recorded several sounds but deleted some of them again because the sound was poor. There are inconsistencies between what the map shows (figure 1) to all participants and me alone. Why this discrepancy?". According to her diary, on January 23, 2020, she experienced poor sound quality:

> Recording offline is the only way that the application works, but that cannot be the point. This time I was on Strandvejen, where I tried to capture the sound of roadworks, machinery and piling with lots of noise. After all, this is precisely one of the goals of the project. The sounds I recorded offline were not transferred when I came online again. I lost four recordings. This problem has been reported to the project team

These examples demonstrate that Louise has struggled with the poor sound quality from the very beginning of the project, including inconsistencies in visualisations of data on the project's interactive map. To improve the sound quality, she experimented with different sound recording practices. E.g. her recordings of fountains in Copenhagen demonstrated that the M system provided inaccurate feedback: "your 'sound robot' [AI system] is not able to recognise splashing and rippling water; it recognises it as a 'rumbling stomach' or 'heavy breathing'".

When Louise downloaded her recordings and listened to them carefully, she discovered that whenever people approached the fountain, the mobile phone's software would reduce the splashing water's sound while amplifying the nearby human voices. The result was a distorted soundscape where neither fountain nor human voices were clear. Louise's detailed descriptions and dialogue with the communication officers contributed to significant changes to the AI systems' data processing and the mobile app's visual feedback. Hence it improved relevant user experience issues concerning the mobile app's general use and understanding.

Communications between participants and project members are crucial to a cs project's continued development since citizens have the potential to uncover a project's shortcomings and create new knowledge about how technology and technological issues influence, constrain, and shape participants' involvement in the project. Citizens' critical work not only influences the CS project and develops it further but also points out how things could be designed differently.

This section demonstrated how citizens' different ways of knowing, doing and directly influencing the materialisation of the CS project. Their entangled material practices of knowing and making visible particular kinds of agencies and possibilities for reconfiguring epistemic oppression (Dotson, 2014), constraints, and injustices that would be missed if we assumed that agency is exclusively human and central to the production of knowledge. The scientific citizen's entanglements with the technoscientific apparatus are dynamic relations of responsibility where nothing is given in advance, apparatuses are not "social formations of power that foreclose agency and produce ideological subjects." (Barad, 2007, p. 240). On the contrary, structures are material-discursive phenomena produced and reconfigured through humans' and nonhumans' intra-active engagements; consequently, apparatuses, including the bodily production of knowledge, and are themselves produced and reworked in iterative ways.

**Theme 3: Entangled responsibility and enactments of concern**

*Enactments of responsibility - potentials for designing otherwise*
Citizens are intra-actively reconfiguring the project through different material engagements with the project. From the early development of the project through interviews and participatory design workshops, citizens enact themselves as citizens and raise their voices and their matters of concern and care in different situations and contexts.

One of my design workshops at Amager Commons, (Amager Fælled), a natural area and public commons in Copenhagen, advanced from the statement: "We need a shared language, a sound community, to create relationships with nature to be free". The concept emerged from the workshop members' concern about the protection and extinction of living species at the Amager Commons (AC), which are under-thread because of the municipality's plans to build houses on the ground. The participants' concern for nature became more explicit during the workshop and through our collective sound explorations, sound mappings, discussions and concept development at the site; through our intra-active material engagements, their passion for protecting nature shaped the focus of the workshop and the outcome of the workshops

One of the participants reflects on the political and ethical possibilities of the SOD project. That could strengthen civic capacities to act, making the public more aware of the natural landscape in Copenhagen. Kristine (activist, member of AC) reflects:

> If more people become aware of the soundscapes out here [at AC], then it could be a legitimate input into the political debate, where right now it is money that is the input [shaping] factor, [...] but only if there's also a part of the political conversation that's about what kind of soundscapes we'd like to have in our city. I often think that within the next 10 years, we will become more aware of the energy landscapes we have. What kind of landscapes do we want? What kind of energy and soundscapes do we want in the city? This is not legitimate to talk about in the political debate today.

This example demonstrates how the public's material engagement and matters of concern can serve as potentials for designing science-citizens relations with the public around community matters of concern, protection of natural areas, and potentially design digital citizen science engagement that serves community needs and interests, including those defined by the scientists. Hence, co-designing another form of science that serves and involves collaboration between public and scientific capacities to act, holds interesting potential for developing more responsible and accountable forms of digital citizen science that combine contributory and democratic citizen engagement (Hetland, 2020). The public clearly has capacities to act and react to others through their material-discursive "acts of citizenship" (Isin and Nielsen, 2008).

These acts are political, ethical and responsible. They are relational acts involving not only humans but also the nonhumans among us. They perform ways of becoming political that potentially strengthen and transform epistemic practices and scientific knowledge production and collectively raise a plurality of different voices to be heard by creating new sites and scales of struggles that could serve to protect AC against extractive industries' intrusive influence of the natural commons. On the other hand, science can empower citizens' voices and strengthen their capacities for making rights claims that build on strong scientific evidence. Together they can constitute citizen science that considers issues of responsibility and accountability and enhance our knowledge of what designs the agency of the already designed and the consequences of designing (Fry, 2020). Citizen's material engagements in citizen science are crucial here because, according to Barad (2007), "knowing is a material practice of engagement as part of the world in its differential becoming". This has obvious ethical consequences for the human-machine relations of which we are just a part and through which potentials and agencies are articulated, constrained or excluded. We are part of the world in its differential becoming.

This is particularly noteworthy because for the citizen scientists the natural environment, science, and sustainability are understood as inseparable aspects of material-discursive engagements. According to Barad, responsibility is a relational enactment where nothing is given in advance. Consequently, relational responsibility offers an opportunity to develop new forms of collaborative and responsible thinking with and from within an unfolding world. This process starts by realising communal agencies (Escobar, 2020) for the continual creation of communities and their connections to their surroundings.

*Sonic engagement and enactments of concern*

The active engagements of the volunteer contributors and their intra-actions with the project's digital platform demonstrate how people are taking responsibility for the relations of which they are a part. Through their ongoing struggles with inadequate sound quality, poor sound recognition and technological issues in their sound recordings, citizens are expressing and enacting their concern and critique of the project.

In a mail to the project, Louise argues that the soundscape produced with the SOD app might be biased because of interferences from the smart phone:

> The mobile recorder is selective – it does not provide a 'true' sound image. Your "sound robot" had difficulties recognising splashing or rippling water; it perceived it as a rolling stomach or panting. Later, I downloaded my audio files and listened to them carefully (she deleted 3 out of 4 because of poor sound quality). In all four cases, I placed myself close to the splashing water and sheltered from the wind, until the sound was potent.

> #2518 Amaliehaven. The fountain in the middle is a powerful fountain with nozzles that practically splash like fire hoses. In the first half of the audio file, the loud splash of water is clear, but about midway through the audio file, the water recording is barely audible because the mobile captures the sound of talking tourists passing by. This substantiates my theory that the mobile is designed to suppress and filter out "background noise". Even though I was standing very close to the splashing water all the time, the 'sound image' is incorrect because human voices represented the background noise. The sound of splashing water was not abruptly muted.

Using this example, Louise's concern can be seen regarding the quality of sound in the project. In addition, it demonstrates how it affects and constrains her engagement with the project and her ability to generate and share epistemic resources of sufficient quality. In her ongoing experiments, she has found that when she records sounds near human voices, the AI system cannot recognise the sound of splashing water. This is because the voice recognition software in her phone is developed to enhance human voices rather than nonhuman sounds like fountains. In consequence, it provides incorrect feedback, diminishing the quality of the recording.

The machinic entanglements of which Louise is part are reconfiguring her material practices and agencies. They enact cuts that limit her possibilities for recording sound of acceptable quality even when she is standing close to a sound source. Thus, she deletes all her recordings from that day, except one. Moreover, Louise's key findings demonstrate that there is a potential risk of generating sound images, data and research that are too biased. This is because the mobile phone is coded to enhance human voices rather than nonhuman sounds such as splashing water. From this perspective, the entire apparatus of entangled intra-actions, therefore, reconfigure how sound is recorded and produced, including the sound quality, the bias encoded in the recorded data, and how people engage with the project or are excluded from engaging. Barad describes the apparatus of production as being reworked through human-nonhuman intra-actions. It is not just about producing products, but also about making subjects and remaking structures (2007, p. 238).

Louise's experiments is an enactment of responsibility and concern, that questions and contests the projects use of smart phones for scientific data collection:

> Collecting sound is a fascinating project, but don't you think mobile phones alone are unsuitable for these sound data collections? Could you consider the possibility that data collectors upload sound recordings with better equipment, so I can use my little Olympus LS-P1, which is half the size of my mobile phone, and then upload the sound afterwards?

For Louise, the poor sound quality and sound recognition in the project are constraining her engagement with the project. For Louise solid sound quality is imperative, hence she deletes many of her recordings in situations where the smartphone interferes and distorts the quality of the recording. She acquired a small microphone to use with her phone to produce better sound for the project, and here she asks for other ways to participate in accommodating the poor sound quality. With her experiments, and by questioning the scientific process, as well as by using microphones to improve the sound quality, she not only expresses her concerns regarding the deficiencies and epistemic exclusions of the project that should be addressed to ensure participation is worthwhile but also suggests other forms of sonic participation to the project.

Louise acquired a small microphone to use with her phone to produce better sound for the project, and here she asks for other ways to participate to accommodate the poor sound quality. With her experiments, and by questioning the scientific process, as well as by using microphones to improve the sound quality, she not only expresses her concerns regarding the deficiencies and epistemic exclusions of the project that should be addressed to ensure participation is worthwhile but also suggests other forms of sonic participation to the project. With her enactments of concern and contestation, as well as her anti-programmatic behaviour that diverges from protocol, she creates tensions within the project (Kasperowski & Hillman, 2018), that contest the scientists' perspectives, opens up unjust constraints and exclusions, and enable alternative modes of sonic engagement.

**Theme 4: Machinic agencies, boundaries and epistemic exclusions**

The following demonstrates how SOD as a technoscientific apparatus of bodily production is constituted iteratively through human and nonhuman intra-actions. Its boundary drawing practices enacts cuts that both enables, constrains, and sometimes excludes specific human agencies, concepts and knowledge practices (see Barad, 2007, p.147). In the following I will demonstrate how some of these boundary drawing practices can result in unwarrantedly epistemic exclusions of citizen scientists.

*Human-machine intra-actions and epistemic exclusion*

By following the active engagements of the volunteer contributors, the technological issues and shortcomings of the project and how they affect citizen agencies is made visible. One of the volunteer contributors, Jens, describes his experiences with the SOD app, and how a lack of feedback from the system affects his experience:

> The sound recording is not working on my iPhone SE. I have tried several times and followed your instructions. When I press the red button, I see a small, round animation that pulses red. The counter remains at 00:00:00:00. I count thirty seconds to see if the screen changes - it does not. I leave your site with no response without knowing if

> anything has happened at your end of the interaction. I have repeated this exercise a few times, even today at project launch. But now I am about to quit. Is there an opportunity to participate in the project and possibly also benefit from SOD's results in the future?

The example demonstrates that participants are not provided proper feedback from the interface to understand their interactions with the app nor if their contributions are uploaded. Jens started the recording, but the timer never started, and there is not feedback on his intra-actions with the app so he cannot comprehend if the sound recording has started, finished or if the project has received his contribution. The missing feedback constraints his epistemic practices and possibilities for contributing to scientific research and for being fully engaged with the projects epistemic ressources.

Louise's descriptions of her sonic explorations in the forest, demonstrate her critical thinking and concern about the AI system's limitations that operate through the project, and show how epistemic practices are constantly reworked:

> I tried to record more sounds with the app. It works best when recording offline. It simply takes way too long with the artificial analysis. I am losing my patience. It is easier to add the description of the sound yourself. However, maybe that means the robot is not getting smarter? There is a conflict between user interests and project interests, and I do not like the long waiting time – after all, the AI cannot provide a reasonable description of the sounds.

This example shows how Louis's engagement with the mobile platform and AI system intra-actively reconfigures her epistemic practices and capacities to act. Louise discovers the app works best offline because of the AI system's time-consuming data processing. She expresses her dissatisfaction and is about to lose her patience with the AI system's slow data analysis and inadequate sound recognition; hence she finds it easier to add sound descriptions manually. The shift from automatic to manual sound description, a reconfiguration of practice, makes her reflect on how it will affect the training of the ML system. The example, furthermore, suggests, that people are concerned about the data quality in the project, because how will her change in practice affect the system? Louise wants to contribute with sounds of good quality, and help to trin the AI system, but the entire apparatus constraints her epistemic possibilities for being fully engaged and to contribute with better data quality.

*Intra-acting with project protocols*

The following example and correspondence with Morten (field recordist) and a communication officer suggest that the project protocols as a structure in the project, not only constrains but unwarrantly excludes certain forms of epistemic agencies:

> **Morten:** I have made some sound recordings in Bulbjerg (with a good ORTF stereo mic system not with a mobile phone). Would you be interested in such material for your project? You can listen here [link to his Soundcloud]
>
> **Communication officer:** Thank you for your contribution and interest. We would like to have your sound recordings, but you will need to use the website www.lyden-af.dk (SOD started as a web app). The website is in Danish, and you need to record outside somewhere in Denmark. You need to stand still and "tag" the sounds you have heard.
>
> **Morten:** Thanks for your mail. I am not interested in using an app on a Smart Phone to record audio since the quality is very poor compared to the high-end mics I use.

As the examples illustrate, the project has an appeal to people who are passionate about sound and recording; therefore, they are willing to contribute using their sound archives and high-end microphones to the project. Since the AI system is trained on low-quality sounds, such as YouTube videos, the project protocols do not permit sound recordings and data collection with personal equipment or sound contributions of high quality, so participation in the project is constrained to sound recordings through the SOD mobile app.

Another volunteer Peter writes: "I do not have a smartphone, but I have a Zoom H5 sound recorder and a PC. It is an unnecessary restriction if only those with smartphones can participate". These examples indicate that people want to be involved. This project has, however, reduced its potential by rejecting more diverse modes of sonic engagement, as well as contributions of high audio quality. A consequence of this could be that participants who wish to share their experiences, sound archives, and knowledge about sounding practices would be excluded.

Several aspects of the project structure, the mobile app, and the artificial intelligence system limit and exclude citizens' opportunities for active engagement, epistemic practices, sharing epistemic resources, and participating in scientific research. Morten understands that smartphone microphones produce unsatisfactory sound quality compared to his high-end ones, which limits his active involvement in the project. Consequently, volunteers cannot contribute according to their interests and protocols. This type of entangled reconfigurations that restricts and excludes epistemic agencies results in anti-programmatic behaviours within a community because of responsible acts of citizenship. Nonetheless, citizens' participation in the project and their expression of concern opens new possibilities for reconfiguring the project and enabling more diverse forms of engagement by allowing them to contribute with their mobile phones, laptops, and high-end microphones. Finally, the need for alternative forms of science-citizen relations to leverage these potentials and create new forms of responsible engagement is crucial, as they are mutually reinforcing.

**Concluding discussion**

This paper's key findings demonstrated how the technoscientific structures of a CS project participate in enacting, enabling and constraining the epistemic agencies of a community of volunteer contributors, the formation of the scientific citizen, and the production of scientific knowledge. Karen Barad's agential realist ontology demonstrated its usefulness as a framework for making distinct possibilities visible for reconfiguring unjust constraints, boundaries, and dynamic power structures that might be missed when agencies are assumed to be predominantly human. I elaborate on these key findings in the following.

Based on my analyses of the citizen science project as a technoscientific apparatus through which material-discursive practices are reconfigured and reconfiguring possibilities for agency through human and nonhuman intra-actions. I demonstrated that citizens are involved in and part of the overall digital development processes that constitute the production of knowledge and phenomena and reconfigure the interactive engagements of volunteer contributors. For the researcher, SOD project team, and the participants, description and observation always happens as an intra-active part of the apparatus; there is no inside nor outside (Barad, 2007, p. 396). I argued that citizens enact citizenship by bringing their attention to issues they care about - that might be missed by the scientists - but also through their direct contributions and reconfigurations of the platform and AI system. I pointed out that the material-discursive enactments always entail human and nonhuman reconfigurations. Thus, agency is iteratively coming into being through the reconfiguration process and the human-nonhuman entanglements.

Through their material-discursive engagements with the project, the scientific citizens bring attention to different matters of concern but also of interest that affects the ongoing development of the CS project. Key findings show that the scientific citizen is interwoven in social and scientific practices that open new sites of negotiation and communication. Furthermore, the analysis points out that citizens take ownership of the CS project and make it their own by drawing attention to issues and agendas they care about and integrating new layers to their engagement with SOD. From this perspective, their intra-actions with the project set a different direction that opened for further collective ways of thinking about sound and sonic engagements that could potentially benefit the ongoing development of the CS project. In addition, I stated that citizens enact citizenship by criticising and contesting the scientific process and by pointing out and making visible the shortcomings in the project, such as poor sound quality and technological and epistemic exclusions. Moreover, we heard that citizens' engagements with the project directly influence the ongoing materialisation of the project through their involvement in the project. While the engagements of citizens suggest how things could be made different, they also point out the importance of designing new sites of citizen science communication.

Through their entanglements with the project, citizens are intra-actively reconfiguring the project, but the project also reconfigures the very bodies and doings of the participants. Key findings suggest that by co-designing with citizens, there are possibilities for designing more responsible, political and ethical forms of citizen science that strengthen new science-citizen relations.

Finally, I demonstrated how the performative effects of the technoscientific apparatus affect the enactments of scientific citizenship and how the scientific citizen is constituted intra-actively through humans and nonhuman entanglements. The citizen scientist's enactments of concern show that people wish to contribute with high data quality and sound quality. However, the project protocols – understood as dynamic power structures – do not allow this. Therefore, the technoscientific apparatus enacts agential cuts and boundaries that both enable and exclude practices and agencies from being co-constituted. From this perspective, I concluded that to design more responsible forms of citizen science engagement; we need more knowledge on the performative effects of the design and how the material and discursive in the design keep designing and enacting unsustainable futures. The first step in this transformation is to acknowledge our relational responsibility for the materialisation process of which we are part because these processes have a role to play in the formation of the scientific citizen and the production of knowledge.

**Acknowledgements**

I am thankful to the people who have shared their thoughts and experiences and been an essential part of the participatory design workshops and sonic explorations. A special thanks to Louise, who contributed detailed information about her experiences in the CS project.

**Data availability**

The author confirm that the data supporting the findings of this study are available within the article [and/or] its supplementary materials.

**Conflict of interest statement**

On behalf of all authors, the corresponding author states that there is no conflict of interest.

Nowotny, H. (2014). Engaging with the political imaginaries of science: Near misses and future targets. *Public Understanding of Science*, *23*(1), 16–20. https://doi.org/10.1177/0963662513476220

Ponti Marisa & Seredko Alena. (2022). Human-machine-learning integration and task allocation in citizen science. *Humanities & Social Sciences Communications*, *9*(1), 1–15. https://doi.org/10.1057/s41599-022-01049-z

Robert V Kozinets. (2015). *Netnography: Redefined* (2. ed.). Sage Pubns Ltd.

Robinson, L. D., Cawthray, J. L., West, S. E., Bonn, A., & Ansine, J. (2018). Ten principles of citizen science. In *Citizen science: Innovation in open science, society and policy* (pp. 27–40). UCL Press.

Schäfer, T., & Kieslinger, B. (2016). Supporting emerging forms of citizen science: A plea for diversity, creativity and social innovation. *Journal of Science Communication*, *15*(2), Y02.

Shirk, J. L., & Bonney, R. (2018). Scientific impacts and innovations of citizen science. In S. Hecker, M. Haklay, A. Bowser, Z. Makuch, J. Vogel, & A. Bonn (Eds.), *Citizen Science* (pp. 41–51). UCL Press; JSTOR. http://www.jstor.org.ep.fjernadgang.kb.dk/stable/j.ctv550cf2.10

Stilgoe, J., Lock, S. J., & Wilsdon, J. (2014). Why should we promote public engagement with science? *Public Understanding of Science (Bristol, England)*, *23*(1), 4–15. https://doi.org/10.1177/0963662513518154

Suchman, L. (2007). *Human-machine reconfigurations*. Cambridge University Press.
29